\documentclass[aps,prl,nofootinbib,amsmath,twocolumn,preprintnumbers]{revtex4-1}
\usepackage{amsmath,amssymb}
\usepackage{graphicx}
\usepackage[export]{adjustbox}

\usepackage{color}
\definecolor{Mahogany}{rgb}{0.62,0.24,0.15}
\definecolor{DarkRed}{rgb}{0.6,0,0}
\definecolor{DarkGreen}{rgb}{0,0.6,0}
\definecolor{DarkBlue}{rgb}{0,0,0.6}
\definecolor{purple}{rgb}{0.6,0,0.8}
\usepackage[colorlinks=true,linktocpage=true,linkcolor=DarkBlue,citecolor=DarkBlue,urlcolor=DarkGreen]{hyperref}

\usepackage[hang,flushmargin]{footmisc}
\usepackage{placeins}
\usepackage{bbold}
\usepackage{multirow}
\usepackage{slashed}
\usepackage{cancel}
\usepackage{xcolor}

\newcommand{\fig}[1]{Fig.~\ref{#1}}
\newcommand{\eq}[1]{Eq.~\eqref{#1}}

\begin{document}

\title{Probing New Gauge Forces with a High-Energy Muon Beam Dump}

\author{Cari Cesarotti}\email{ccesarotti@g.harvard.edu}
\author{Samuel Homiller}\email{shomiller@g.harvard.edu}
\author{Rashmish K.~Mishra}\email{rashmishmishra@fas.harvard.edu}
\author{Matthew Reece}\email{mreece@g.harvard.edu}
\affiliation{\vspace{0.5em}Department of Physics, Harvard University, Cambridge, MA, 02138\\}


\begin{abstract}
We propose a new beam dump experiment at a future TeV-scale muon collider. A beam dump would be an economical and effective way to increase the discovery potential of the collider complex in a complementary regime. In this work we consider vector models such as the dark photon and $L_\mu-L_\tau$ gauge boson as new physics candidates and explore which novel regions of parameter space can be probed with a muon beam dump. We find that for the dark photon model, we gain sensitivity in the moderate mass (MeV--GeV) range at both higher and lower couplings compared to existing and proposed experiments, and gain access to previously untouched areas of parameter space of the $L_\mu-L_\tau$ model.
\end{abstract}

\maketitle

\section{Introduction}
\label{sec:intro}

The Standard Model (SM) of particle physics has proven to be remarkably successful. However, there is an abundance of empirical evidence that it is incomplete. While there are a variety of channels we can explore to discover new physics, colliders provide a clean and controlled experimental environment to identify the particle content of beyond the SM phenomena. 

As we look to advance the energy and intensity frontier of collider physics, a possibility with growing interest is the construction of a TeV-scale $\mu^+ \mu^-$ collider \cite{Antonelli:2015nla,Long:2020wfp,MICE:2019jkl,Delahaye:2019omf,Neuffer:1983jr,Delahaye:2013jla}. Such a muon collider ($\mu C$) is a particularly compelling option as it affords a complementary physics program to that of a high-energy hadron collider like the LHC.
For example, with a $\mu C$ we gain access to direct couplings of both electroweak-mediated and second-generation processes~\cite{Chen:2016wkt, Buttazzo:2018qqp, Chiesa:2020awd, Costantini:2020stv, Han:2020uid, Han:2020pif, Han:2020uak, Han:2021kes, AlAli:2021let, Asadi:2021gah, Ruiz:2021tdt}.
Additionally, with increased available center-of-mass energy, we can expand our discovery prospects for massive new physics. 

Since the cost of a $\mu C$---or any future high-energy collider---is substantial, it is prudent to consider possible auxiliary experiments that extend the physics program of the collider facility. An economical extension with remarkable and complementary discovery potential is a beam dump. In a beam-dump experiment, the high energy muon beam is `dumped' into a dense material to greatly increase the total rate of interaction at the price of center-of-mass energy. This experimental setup can therefore test couplings too small to be probed at the main collider by several orders of magnitude in a slightly lower mass range \cite{Bjorken:2009mm,Essig:2010gu, Kanemura:2015cxa, Chen:2017awl, Sakaki:2020mqb,Asai:2021xtg,Sieber:2021fue}.

In this letter we propose the construction of a beam-dump experiment to be included in the design of a future $\mu C$. We consider benchmark models of moderate-mass, weakly coupled new vector particles that are inaccessible at any other terrestrial experiment. First, we consider the dark photon scenario~\cite{Holdom:1985ag, Pospelov:2007mp, Arkani-Hamed:2008hhe, Pospelov:2008zw}, for which similar past proposals have focused on electron beams; see, e.g.,~\cite{Reece:2009un, Bjorken:2009mm, Essig:2010xa}. Here, the main novelty of a proposed $\mu C$ is high energy, which can provide access to  a different range of masses and (due to the large boost) lifetimes than lower-energy electron beam dumps. We also consider a model for which a muon beam is uniquely well-suited, namely the gauged flavor symmetry $L_\mu-L_\tau$~\cite{Foot:1990mn, He:1990pn,  Ma:2001md,Huang:2021nkl}. In this case, the gauge boson is produced much more copiously  from a muon beam than an electron or proton beam. 
We present the projected reach of such an experiment for several generic experimental configurations.

\section{Production from a High Energy Muon Beam}
\label{sec:prodMuBeam}
\subsection{Preliminaries}

In what follows we restrict our attention to the new physics scenario of a new vector particle, which we generically call $Z'$, coupling to a current including muons. The effective Lagrangian of interest for a new $U(1)'$ gauge boson is
\begin{multline}
\mathcal{L} \supset \mathcal{L}_\text{SM}  -\frac{1}{4}Z'_{\mu\nu}Z'^{\mu\nu} + \frac{1}{2} m_{Z'}^2 Z'^\mu Z'_\mu
- \\ \sum_{l\in e, \mu, \tau}\left( i g Q_l \bar{l} \gamma^\mu Z'_\mu l  + i g Q'_l \nu_l^\dagger \bar{\sigma}^\mu Z'_\mu \nu_l \right).
\label{eq:fullLag}
\end{multline}
We consider two $U(1)$ models: dark photons $(Q_l =1, Q_l' =0, g = \epsilon e)$ and the gauged flavor symmetry $L_\mu-L_\tau$ $(Q_{\mu, \tau} = Q_{\mu, \tau}' = \pm1)$. In the latter case, kinetic mixing will also be present (at least through loops of muons and taus; see, e.g.,~\cite{Bauer:2018onh}), but the effect is both small and dependent on details of the UV completion, so we will neglect it in our discussion.

The photon--dark photon interaction is often defined via kinetic mixing in a basis of quantized charges, but the above formulation can be derived through the field redefinition $A_\mu \rightarrow A_\mu + \epsilon Z'_\mu$ to generate a coupling to the EM current. Both the dark photon and gauged flavor symmetry models are of interest at a high-energy muon collider, although they are differently motivated. Dark photons can be produced directly at any charged particle collider, but the extended reach of our proposed experiment comes from the increased center-of-mass energy. 
The $L_\mu-L_\tau$ model, on the other hand, would benefit uniquely from a muon collider, as this would be the first experiment capable of direct production at high energies. 
\begin{figure}[h!]
\centering
\includegraphics[width=0.4\textwidth]{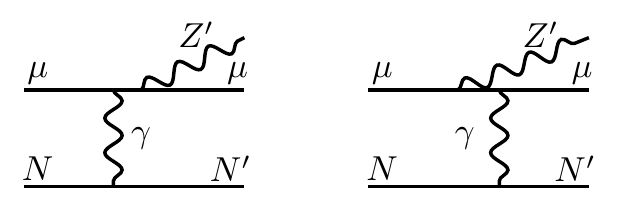}
\includegraphics[width=0.4\textwidth]{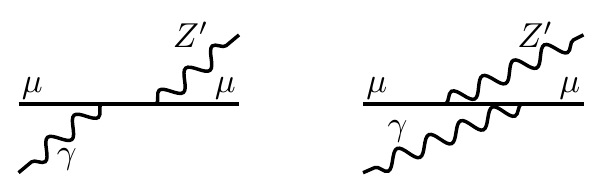}
\caption{Top: the dominant bremsstrahlung production for a vector particle $Z'$ at a muon beam dump. Bottom: the same production process in the Weizs\"acker-Williams approximation.}
\label{fig:WW}
\end{figure} 
\subsection{Cross Section}
The dominant production mechanism is the $2\rightarrow3$ bremsstrahlung process shown in the top of \fig{fig:WW}, where the incoming high-energy muon exchanges a virtual photon with a nucleon in the target and radiates a $Z'$~\cite{Kim:1973he, Tsai:1973py, Tsai:1986tx, Bjorken:2009mm}.
To compute this cross section, we use the Weizs\"acker-Williams approximation~\cite{vonWeizsacker:1934nji, Williams:1934ad}. For relativistic incoming muons, the exchanged photon is nearly on-shell. We can therefore approximate the full scattering process $\left(\mu (p) + N(P_i) \rightarrow \mu (p') + N'(P_f)+ Z'(k)\right)$ with the $2 \rightarrow2$ process $\left(\mu (p) + \gamma(q) \rightarrow \mu (p') + Z'(k)\right)$ shown in the bottom of \fig{fig:WW}, evaluated at minimum virtuality $t_\text{min} \equiv -q^2_\text{min}$, weighted by the effective photon flux. The cross section in the lab frame is 
\begin{multline}
\frac{d\sigma(p + P_i \rightarrow p' + k + P_f)}{dE_{Z'}d \cos \theta_{Z'}} = \left( \frac{\alpha_\textsc{EM} \chi}{\pi}\right) \left( \frac{x E_0 \beta_{Z'}}{(1-x)} \right)\\ \times \frac{d \sigma (p+q \rightarrow p' + k)}{d(p \cdot k)}\bigg \rvert_{t=t_\text{min}},
\label{eq:fullWW}
\end{multline}
where $E_0$ is the energy of the incoming muon beam, $x\equiv E_{Z'}/E_0$ is the fraction of energy of the $Z'$, $\theta_{Z'}$ is the angle of emission, and $\beta_{Z'} \equiv \sqrt{1-m_{Z'}^2/E_{Z'}^2}$. The effective photon flux is parameterized by $\chi$, defined as 
\begin{equation}
\chi \equiv \int_{t_\text{min}}^{t_\text{max}} dt\, \frac{t-t_\text{min}}{t} G_2(t),
\end{equation}
where $G_2(t)$ is the electric form factor of the target atom, including both atomic and nuclear and elastic and inelastic effects, following the approximation in~\cite{Bjorken:2009mm}. 

\begin{figure}[h!]
\includegraphics[width=0.4\textwidth]{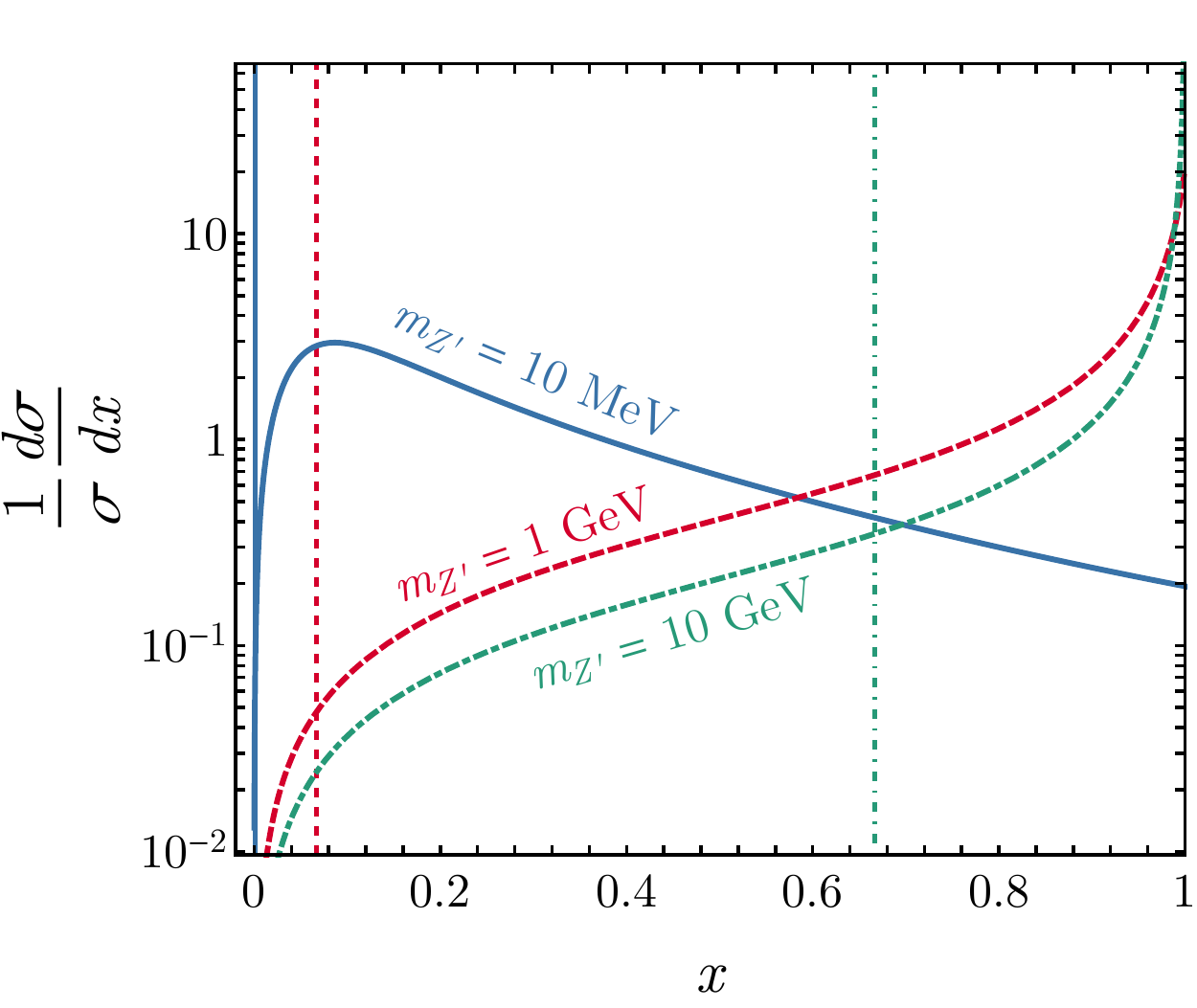}
\caption{Distributions of the fraction of the beam energy $x$ carried away in the emission of the $Z'$ for various masses. The vertical lines indicate where the boost factor $\gamma=100$ for the corresponding mass, thus for all masses the majority of emission occurs to the right of these lines in the highly relativistic regime.}
\label{fig:sigmaDist}
\end{figure}

This approximation scheme is valid in the regime of highly relativistic beam particles and emitted vector particles:
\begin{equation}
\frac{m_\mu}{E_0}, \ \ \frac{m_{Z'}}{xE_0}, \ \ \theta_{Z'} \ll 1.
\end{equation}
After integrating out the angular dependence, the differential cross section in $x$ is 
\begin{equation}
\frac{d\sigma(2\rightarrow3)}{dx} = \frac{8\alpha_\textsc{EM}^2 \alpha_g Q_\mu^2 \chi  \beta_{Z'}}{m_{Z'}^2 \frac{1-x}{x} + m_\mu^2 x} \left( 1 - x + \frac{x^2}{3}\right),
\label{eq:cs2to3}
\end{equation}
where $\alpha_g \equiv \frac{g^2}{4\pi}$. The normalized distribution of \eq{eq:cs2to3} is plotted in \fig{fig:sigmaDist} for various vector masses $m_{Z'}$. For all values of mass we find that the probability of emission has support primarily in the highly relativistic regime. 

Note that in the context of an electron beam dump experiment, the term proportional to the beam particle mass in the denominator of \eq{eq:cs2to3} can be neglected.
Here, on the other hand, we are interested in $m_{Z'}$ near the muon mass, so we cannot drop this term.  
\subsection{Signature}
The signal of interest for this experimental setup is a dilepton final state ($e^+e^-$ or $\mu^+\mu^-$). From the Lagrangian defined in \eq{eq:fullLag}, the decay rate to massive leptons is 
\begin{equation}
\Gamma\left( Z' \rightarrow l^+ l^-\right) = \frac{g^2 Q_l^2}{12\pi} m_{Z'}\left( 1 + \frac{2m_l^2}{m_{Z'}^2} \right)\sqrt{1 - \frac{4 m_l^2}{m_{Z'}^2}}.
\label{eq:zPrimeWid}
\end{equation}
We restrict our attention to this signature for ease of detection prospects. 

Before we proceed with the details, we can check that a multi-TeV muon collider provides the right environment to extend the search boundaries of these models. Taking the dark photon as a benchmark, the approximate decay length (estimated from \eq{eq:zPrimeWid}) 
\begin{multline}
l_{Z'} \equiv c \gamma \tau_{Z'} \approx x \left( \frac{E_\text{0}}{\text{TeV}} \right)  \left( \frac{\text{GeV}}{m_{Z'}} \right)^2 \left( \frac{10^{-7}}{\epsilon} \right)^2 \times 10 \text{ m},
\label{eq:cgammatau}
\end{multline}
where $x \sim 1$.
This suggests that with a modest-size experiment, a TeV beam dump can dramatically expand the reach of these vector models. 

\subsection{Number of Signal Events}
The differential number of signal $Z'$ events per energy fraction $x$ and position $z$ along the beamline is given by the equation
\begin{multline}
\frac{dN}{dx\, dz} = N_\mu \frac{X_0}{m_T} \int_{E_{Z'}}^{E_0} dE_1 \int_0^T dt\, I(E_1; E_0, t) 
\\ \times \left( \frac{E_0}{E_1} \frac{d\sigma}{dx'}\right)_{x' = \frac{E_{Z'}}{E_1}} \frac{dP(z-\frac{X_0}{\rho}t)}{dz},
\label{eq:mastEq}
\end{multline}
where $N_\mu$ is the number of incident muons on target, $m_T$ is the mass of a target atom, and $X_0$ and $\rho$ are the unit radiation length and density of the target respectively.\footnote{In the literature, $1/m_T$ is often written as $N_0 / A$ where $N_0$ is Avogadro's number and $A$ is the atomic number of the target. This assumes that $X_0$ is measured in particular units, g/cm$^2$.}
We parameterize the position along the length of the target in terms of the dimensionless parameter $t$ from $0$ to $T$, such that the full length of the target is $L_\text{tar} = \frac{X_0}{\rho}T$. 

The energy of the beam particle after radiative losses in the material is modeled by the function $I(E_1; E_0, t)$. However, the effective radiation length of a muon in reasonable target materials (e.g., water or lead) is 50~m to 1~km~\cite{Zyla:2020zbs}. For the proposed experimental setup, we can safely assume zero radiative losses and replace 
\begin{equation}
I(E_1; E_0, t) = \delta(E_1 - E_0). 
\end{equation}

\begin{figure}[h!]
\includegraphics[width=0.4\textwidth]{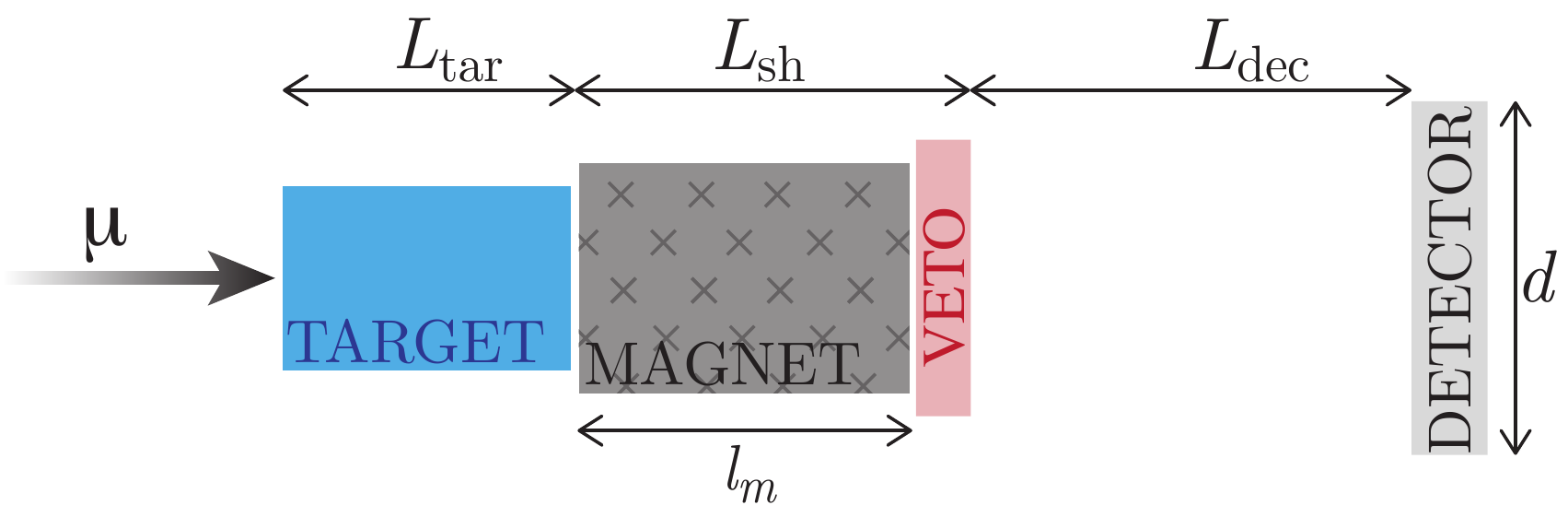}
\caption{Schematic of a muon beam dump experiment. The lengths are not drawn to scale. }
\label{fig:expSetup}
\end{figure}

The differential probability of $Z'$ decay is given by 
\begin{equation}
\frac{dP(z)}{dz} = \frac{1}{l_{Z'}} e^{-z/l_{Z'}}.
\end{equation}
As indicated in \eq{eq:cgammatau}, the decay length $l_{Z'}$ is a function of the $Z'$ mass $m_{Z'}$ and energy $x E_0$. 
After integrating over the muon beam energy and target thickness, \eq{eq:mastEq} simplifies to 
\begin{multline}
\frac{dN}{dx} = N_\mu \frac{\rho\, l_{Z'}}{m_T} \frac{d\sigma}{dx} \times \\
\left( e^{L_\text{tar}/l_{Z'}} - 1\right)e^{-(L_\text{tar}+L_\text{sh})/l_{Z'}} \left(1 - e^{-L_\text{dec}/l_{Z'}} \right).
\label{eq:masterEq}
\end{multline}
where the various length scales are illustrated in \fig{fig:expSetup}.

%
%
\section{Experimental Setup}
\label{sec:expSetup}
The experimental setup of a beam dump experiment is shown in \fig{fig:expSetup}. The high energy muon beam is dumped into a solid material target of length $L_\text{tar}$. Immediately after the target is a shield of length $L_\textrm{sh}$ dedicated to removing the residual beam and any background. This includes both a region with a strong magnetic field of length $l_m$ and a veto mechanism to identify or remove any remaining particles. Beyond the shielding is the fiducial decay region of length $L_\text{dec}$ before the detector. To be detected, signal events must be produced in the target and decay in fiducial region. 

From \eq{eq:masterEq} it is clear that the length scales of the various components of the experiment have dramatic impact on the number of signal events observed. New particles that are produced too early get shielded or vetoed and are missed, but particles that live too long escape the experimental hall altogether. 

For concreteness, we choose experimental parameters such as length scales at reasonable orders of magnitude. The purpose of this letter is to demonstrate the feasibility of beam dump experiments at future muon colliders. Since we cannot know the available technology or experimental design constraints at this point, we do not attempt to optimize the experimental design.
\subsection{Target}
In this work we consider a target made of water at standard temperature and pressure, with a length of $L_\text{tar} = 10$ m. Other economical and common choices for a beam dump target include pressurized water, lead or tungsten. Higher density materials will increase the cross section as the effective photon flux increases, but may be too expensive or infeasible to install at the beam dump site. 
\subsection{Shielding}
Since the $Z'$ and its subsequent decay products are highly boosted, it is imperative to remove the central muon beam for a near zero-background signal. To do so, the shielding region must remove the muon beam from the geometric acceptance of the detector. Since high-energy muons are extremely difficult to absorb, we instead propose to deflect the beam using a strong magnetic field. 

The size of the detector is set by the opening angle of the dilepton decay products of the boosted $Z^{\prime}$. The decay products are produced extremely forward with an emission angle in the lab frame of $\theta_\textrm{max} \lesssim m_{Z'}/xE_0$. Note that, in the parameter space of  interest where $m_{Z'} > m_\mu$, the maximum angle of $Z'$ emission $\theta^{Z'}_\textrm{max} \sim \frac{\sqrt{m_{Z'} m_\mu}}{E_0}$ is parametrically smaller and therefore negligible. 

The shielding magnet must be sufficiently strong to divert a roughly TeV-scale muon beam at a greater angle $\theta_\text{mag} > \theta_\textrm{max}$. If we assume the magnetic field is constant over a length $l_m$ just after the target, the field strength $B$ must be approximately 
\begin{equation}
\frac{B}{\mathrm{Tesla}} \sim \theta_\text{max} \times \frac{E_0}{\text{GeV}} \times \frac{\text{meter}}{l_m}.
\end{equation}
For the parameter space of interest, this corresponds to sub-Tesla magnetic fields over a few meters, which is comparable to or smaller than similar vetos proposed in other future beam-on-target experiments \cite{SHiP:2015vad}. 

Since bending a high-energy beam of charged particles will emit some radiation, there must also be a passive veto beyond the magnetic field. This veto can be made of a combination of dense material to absorb photons, such as iron or lead, as well as a charged particle tracker to catch any charged remnants. 

We assume that the total length of this shielding region will be contained in $L_\text{sh} = 10$ m, but we discuss the consequences of a larger $L_\text{sh}$ later with the results. 
\subsection{Detector}
Once a $Z'$ has been produced in the target and propagates beyond the shielding region, it must decay in the fiducial region before the detector in order to be observed. As previously mentioned, our signal of interest is a dilepton resonance ($e^+e^-$ or $\mu^+\mu^-$). This signal can be observed with a relatively simple detector setup: a charged particle and muon tracker. Again, as the state-of-the-art detection technology is not yet known, we do not provide a detailed description of the system. The size of the detector $d$ must be roughly 
\begin{equation}
d \sim \theta_\text{max} L_\text{dec}.
\label{eq:detectorSize}
\end{equation}

From \eq{eq:masterEq}, it is clear that a longer experimental hall increases the number of $Z'$ particles that decay in acceptance. As a reasonable benchmark we set $L_\text{dec} = 100$ m, which corresponds to a detector size of $d\lesssim$ 2 m for the relevant range of masses. Note that for this exploratory study, we consider a minimally instrumented scenario. However, one could envision more sophisticated detection scenarios that involve instrumenting along the fiducial region and accounting for missing energy signals as well. 

\section{Reach for New Gauge Forces}
\label{sec:reach}
We present the reach of both a dark photon and $L_\mu - L_\tau$ $Z'$ model. For concreteness we consider the reach with a 1.5 TeV beam (corresponding to a 3 TeV collider), a standard benchmark in $\mu C$ literature \cite{Delahaye:2019omf}. 

For the dark photon scenario, we show the existing constraints from $e^+ e^-$ or $\mu^+\mu^-$ resonance searches at BaBar~\cite{BaBar:2014zli}, NA48~\cite{Na482:2015wmo}, the A1 Experiment at the Mainz Microtron~\cite{Merkel:2014avp}, KLOE~\cite{KLOE-2:2011hhj, KLOE-2:2012lii, KLOE-2:2014qxg, KLOE-2:2016ydq},  and LHCb~\cite{LHCb:2019vmc}; previous beam dump experiments, such as E141~\cite{Riordan:1987aw} and E137~\cite{Bjorken:1988as, Batell:2014mga, Marsicano:2018krp} at SLAC, E774 at Fermilab~\cite{Bross:1989mp}, CHARM~\cite{CHARM:1985anb, Gninenko:2012eq} and NuCal~\cite{Blumlein:1990ay, Blumlein:2011mv, Blumlein:2013cua}; as well as constraints from Supernova 1987A~\cite{Chang:2016ntp} in gray. 
Additionally we plot the projected reach from other future experiments including Belle-II~\cite{Belle-II:2018jsg}, LHCb~\cite{Ilten:2015hya, Ilten:2016tkc}, SHiP~\cite{Alekhin:2015byh}, and AWAKE~\cite{Caldwell:2018atq}.

\begin{figure}[h!]
\centering
\includegraphics[width=0.45\textwidth, left]{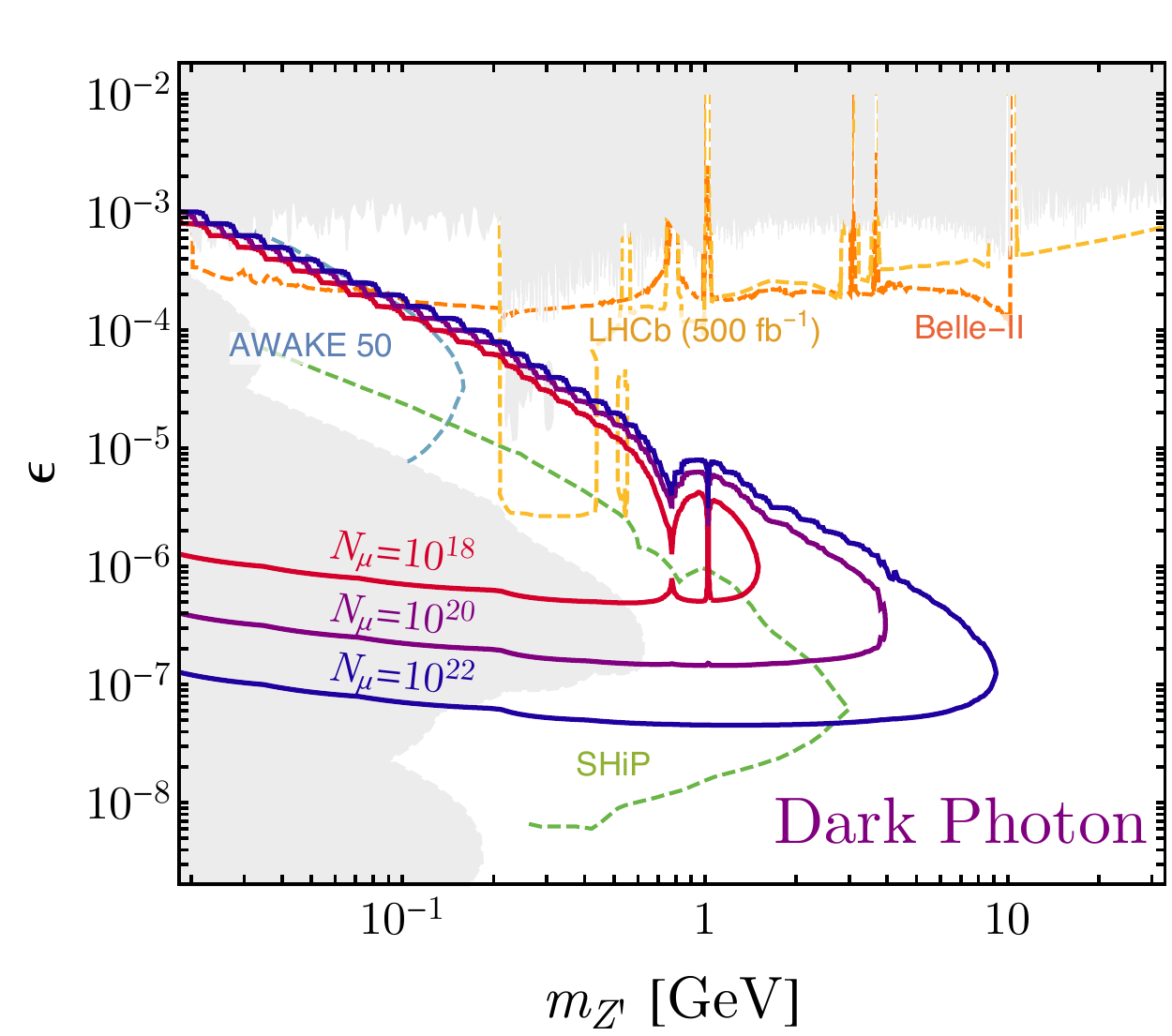}
\caption{Contour plots indicating 5 signal events detected with $N_\mu =10^{18}, \ 10^{20}, 10^{22}$ with a beam energy $E_0 = 1.5$ TeV. The dips in the contours near $m_{Z'} = 1$ GeV occur when there is resonant production in the $Z' \rightarrow$ hadron decay channel, thus reducing the dilepton branching ratio.}
\label{fig:dpReach}
\end{figure}

The projected sensitivity of the $\mu C$ beam dump to a dark photon $Z'$ is shown in \fig{fig:dpReach}. Since the number of muons delivered to target cannot be known at this stage, we provide three reach curves reflecting conservative to optimistic projections of $N_\mu$. A discussion of these choices can be found in the appendix. 
The experiment would expand the reach in parameter space not only beyond existing constraints, but also in complementary regions to other future experiments. The gain in coverage occurs mainly in the directions of larger coupling and mass. This is due to the high energy of the beam, and therefore production of highly boosted $Z'$ particles.

The boundary of the discovery region at large couplings occurs when the $Z'$ decays too early and is vetoed. However, a relativistic $Z'$ will live longer in the lab frame and therefore can decay in the fiducial region. At this unprecedented beam energy, $Z'$s at higher masses than before are sufficiently boosted to survive past the veto. If the shielding length $L_\text{sh}$ is extended, then sensitivity degrades in the large coupling regime while leaving the bottom edge unmoved. However, if the beam energy $E_0$ is increased, both the upper and lower boundaries are shifted upwards to higher couplings.  
\begin{figure}[h!]
\centering
\includegraphics[width=0.45\textwidth, left]{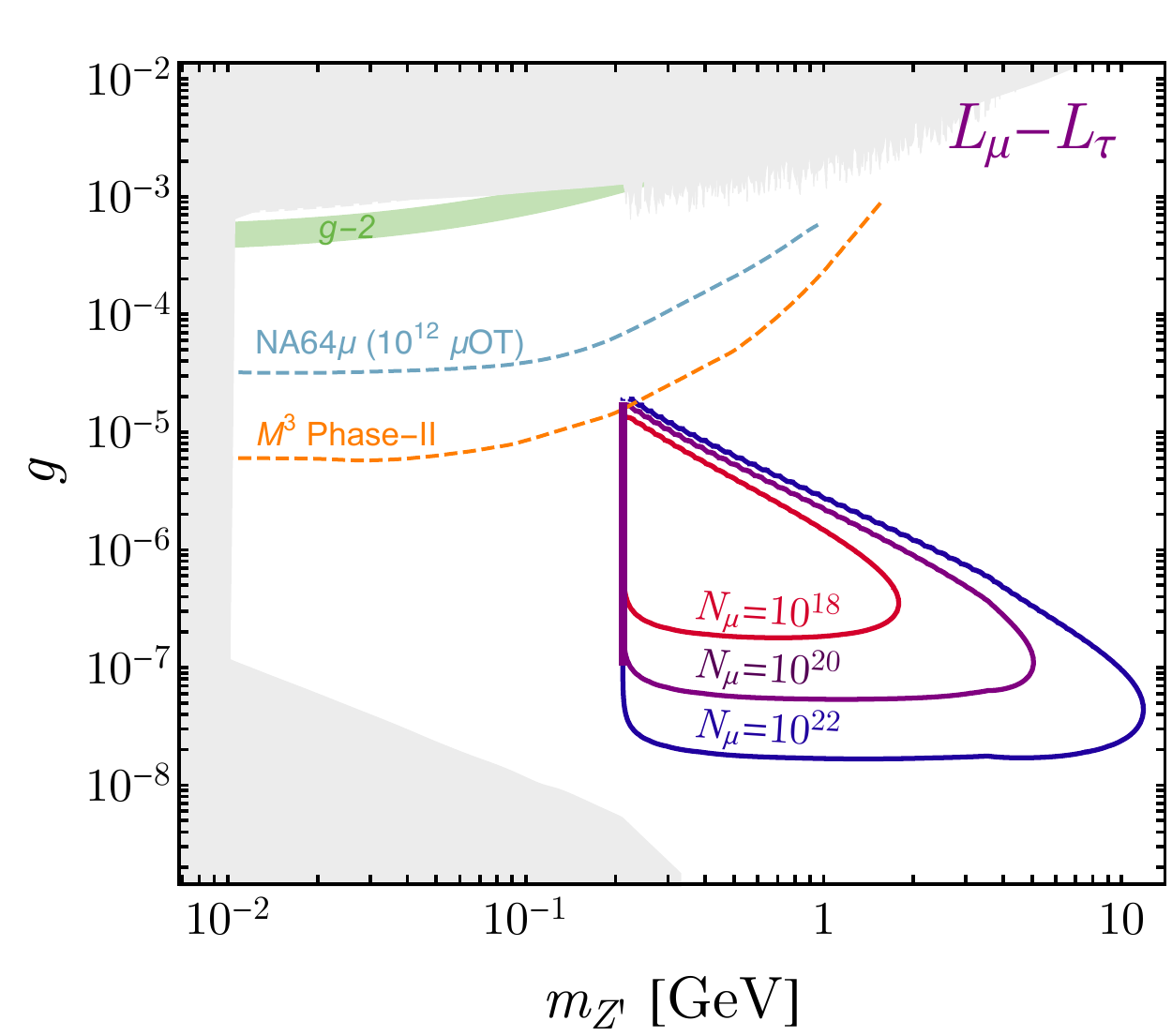}
\caption{Same as \fig{fig:dpReach} but for the $L_\mu-L_\tau$ model. The sensitivity is bounded at lower $m_{Z'}$ by the dimuon production threshold, since the electron channel is not open in this model.}
\label{fig:lmultauReach}
\end{figure}

The parameter space of the $L_{\mu} - L_{\tau}$ model is notably unconstrained in the region $g \lesssim 10^{-3}$ and $m_{Z'} \gtrsim10$ MeV.
Existing constraints come from measurements of the primordial abundances of light nuclei~\cite{Escudero:2019gzq}, from observations of SN1987A \cite{Croon:2020lrf},
from measurements of the anomalous magnetic moment of the muon,\footnote{Given the current discrepancy between the theoretical prediction~\cite{Aoyama:2020ynm} and experimental measurement of $(g-2)_{\mu}$~\cite{Muong-2:2021ojo}, we take the $5\sigma$ upper limit as a constraint, and show the $2\sigma$ preferred region in green in \fig{fig:lmultauReach}.} limits from neutrino trident production~\cite{Altmannshofer:2014pba, CCFR:1991lpl} and searches for $e^+e^- \to \mu^+\mu^- Z'(\mu^+\mu^-)$ at BaBar~\cite{BaBar:2016sci}.
The current bounds are shown in grey in \fig{fig:lmultauReach}. We also show the projected limits from other muon beam experiments, M$^3$~\cite{Kahn:2018cqs} and NA64$\mu$~\cite{Sieber:2021fue}.
Note that other proposed experiments such as Ref.~\cite{Krnjaic:2019rsv} might have comparable reach to NA64$\mu$ and M$^3$.
The reach plot, drawn with the same values of $N_\mu$ and beam energy $E_0$, is shown in \fig{fig:lmultauReach}. The coverage is completely separated from other constraints on this model due to the novelty of both the beam of muons and the energy of the beam.

\section{Conclusions and Future Work}
\label{sec:conc}
A future multi-TeV muon beam dump experiment would provide a window into previously unexplored parameter space for a variety of motivated new physics models. The sensitivity improvements beyond other similar proposed experiments, such as Refs.~\cite{Bjorken:2009mm, Essig:2010gu, Kanemura:2015cxa, Chen:2017awl, Sakaki:2020mqb,Asai:2021xtg,Sieber:2021fue}, stem from two features of such an experimental setup: the increased beam energy and direct coupling to muons. 

 As discussed, the unprecedented beam energy translates to boosted new particles with extended lifetimes. Therefore couplings that would otherwise be too large to be detected at previous beam dump experiments would be accessible. Additionally, a muon collider is uniquely well suited to study models with couplings to muons. In this letter we computed the reach of the gauged $L_{\mu} - L_{\tau}$ symmetry as a motivated example, but this broadly applies to more general dark sectors with non-universal fermion interactions.
 
In the proposed detection strategy, we've taken a minimalist approach to instrumentation. However, in the event of an observed resonance, additional detectors to identify the rate into taus could be used to determine the underlying theory.
If this experiment were to confirm the existence of a gauged flavor symmetry, this would be significant for several areas of particle physics. A gauged $L_\mu - L_\tau$ symmetry could explain the near-maximal mixing between muon neutrinos and tau neutrinos~\cite{Ma:2001md}. It has recently been observed that an $SU(3)$ extension of this group can give rise to a complete model of lepton masses~\cite{Alonso-Alvarez:2021ktn}.
Finally, we\footnote{One of us, anyway.} cannot resist noting that every measured gauge coupling to date is an $O(1)$ number, while \fig{fig:lmultauReach} shows that a discovery of $L_\mu - L_\tau$ at the beam dump would necessarily imply a tiny gauge coupling $\lesssim 10^{-5}$, which would become a powerful constraint on UV physics~\cite{Arkani-Hamed:2006emk}.

While we have focused on models with new vectors, the improved reach would be similarly impressive for other dark sectors with non-vector mediators. Additional searches should include models with new muon-philic scalars or pseudoscalars such as axions. These particles could be produced either directly from the muon or from photon fusion by effective interactions. A detailed study of the reach for these models will be presented in a future publication.

In the current era of particle physics, no stone can be left unturned when searching for new phenomena. A future muon collider and a corresponding beam dump experiment would greatly enhance our sensitivity in novel and complementary regimes to our current and past experimental program.

\section*{Acknowledgments}
We are grateful to Cliff Cheung, Matheus Hostert, Simon Knapen, Johannes K. L. Michel, Cristina Mondino, Clara Murgui, Simone Pagan Griso, Matthew Strassler, and Jesse Thaler for useful discussions. 
CC, SH and MR are supported by the DOE Grant DESC0013607. CC is also supported by an NSF Graduate Research Fellowship Grant DGE1745303.
SH and MR are also supported in part by the Alfred P. Sloan Foundation Grant No.~G-2019-12504.
RKM is supported by the National Science Foundation under Grant No.~NSF PHY-1748958 and NSF PHY-1915071.
MR is also supported by the NASA Grant 80NSSC20K0506.


\section{Notes on the Expected Number of Muons on Target}
\label{app:luminosity}

The number of muons available at a beam dump experiment associated with a high-energy collider depends significantly on the beam design, for which at this time there are only rough proposals. 
Parametrically, the number of muons in a collider beam, per unit time, is given by
\begin{equation}
{\dot N}_{\mu} = f_0 \times n_b \times N_{\mu / \textrm{bunch}}
\end{equation}
where $f_0$ is the muon source repetition rate (i.e., the rate at which the muons are generated by the source, which could be either a proton or a positron beam), $n_b$ is the number of colliding bunches in each beam, and $N_{\mu /\textrm{bunch}}$ is the number of muons per bunch.
Assuming the MAP design parameters~\cite{Delahaye:2013jla, Neuffer:2018yof} ($f_0 = 5\,\textrm{Hz}$, $N_{\mu / \textrm{bunch}} = 2 \times 10^{12}$, $n_b = 1$), this translates to ${\dot N}_{\mu} \sim 10^{20} / \textrm{year}$ generated for collisions.
The actual number of muons available at the beam dump target of course will be smaller, and will depend on how frequently the beam is dumped.
To our knowledge, this frequency is not discussed in any existing design proposals for a muon collider, and will have to be optimized to maximize the number of high-energy collisions while mitigating neutrino radiation and maintaining a high quality beam. Assuming, however, that the muons are dumped after approximately a muon lifetime, $N_{\mu} \sim 10^{20}$ in a multi-year run may be achievable.

While the MAP proposal envisions a proton driver as the source of the muon beam, an alternative possibility is to use a positron source, as in the LEMC scenario. A positron source would have a much smaller number of muons per bunch, but compensates with a much higher repetition rate, $f_0$. The current proposals obtain much smaller luminosities at a high-energy collider, and correspondingly lower number of muons on target for an associated beam dump ($N_{\mu} \lesssim 10^{18}$ assuming the LEMC parameters in ref.~\cite{Neuffer:2018yof}), but future technological advances may significantly increase the muon yield.

\bibliographystyle{utphys}
\bibliography{mudump}

\end{document}